\documentclass[journal,draftcls,onecolumn,12pt,twoside]{IEEEtran}
\usepackage{mathrsfs}
\usepackage{amsfonts}
\usepackage{graphicx,cite,epsfig,amssymb,amsmath}
\usepackage{color,xcolor}
\usepackage{pifont}
\usepackage{stmaryrd}
\usepackage{setspace}
\usepackage{subfigure}
\usepackage{cite}
\usepackage{array}
\usepackage{float}
\usepackage{multirow}
\usepackage{algorithmic,algorithm}
\usepackage{epstopdf}
\usepackage{mathtools}

\floatstyle{ruled}
\providecommand{\tabularnewline}{\\}
\providecommand{\algorithmname}{Algorithm}
\floatname{algorithm}{\protect\algorithmname}
\newcommand{\bm}[1]{\mbox{\boldmath{$#1$}}}
\newtheorem{thm}{Theorem}
\newtheorem{rem}{Remark}
\newtheorem{lem}{Lemma}
\newtheorem{prob}{Problem}

\begin{document}

\title{Latency Optimization for Resource Allocation in Mobile-Edge Computation Offloading}
\author{\IEEEauthorblockN{Jinke Ren, Guanding Yu, Yunlong Cai, and Yinghui He}\\
\vspace*{0.3cm}
\small{\IEEEauthorblockA{College of Information Science and Electronic Engineering \\Zhejiang University, Hangzhou 310027, China\\
                     E-mail: \{renjinke, yuguanding, ylcai, 2014hyh\}@zju.edu.cn}}}
\maketitle

\begin{abstract}
By offloading intensive computation tasks to the edge cloud located at the cellular base stations, mobile-edge computation offloading (MECO) has been regarded as a promising means to accomplish the ambitious millisecond-scale end-to-end latency requirement of the fifth-generation networks. In this paper, we investigate the latency-minimization problem in a multi-user time-division multiple access MECO system with joint communication and computation resource allocation. Three different computation models are studied, i.e.,  local compression, edge cloud compression, and partial compression offloading. First, closed-form expressions of optimal resource allocation and minimum system delay for both local and edge cloud compression models are derived. Then, for the partial compression offloading model, we formulate a piecewise optimization problem and prove that the optimal data segmentation strategy has a piecewise structure. Based on this result, an optimal joint communication and computation resource allocation algorithm is developed. To gain more insights, we also analyze a specific scenario where communication resource is adequate while computation resource is limited. In this special case, the closed-form solution of the piecewise optimization problem can be derived. Our proposed algorithms are finally verified by numerical results, which show that the novel partial compression offloading model can significantly reduce the end-to-end latency.
\end{abstract}


\section{Introduction}
Over the past few years, the explosive popularity of mobile devices, such as smart-phones, tablets, and wearable devices, has been accelerating the development of the Internet of Things (IoT) \cite{IoT_development,Fog_and_IoT}. According to the prediction by Cisco, nearly 50 billion IoT devices will be connected to the Internet by 2020, most of which have limited resources for communication, computation, and storage \cite{IoT_devices}. Due to the exponential growth of mobile data traffic, merely relying on traditional cloud computing is not adequate to realize this ambitious millisecond-scale latency for communication and computation in 5G networks. To keep up with this persistent demand and improve the quality of experience (QoE) for users, the emerging technology of mobile edge computing (MEC) has been gaining significant attention from both academia and industry.

MEC offers application developers and content providers cloud-computing capabilities at the very edge of the mobile network by implementing MEC servers at cellular base stations (BSs), which is also referred to as \emph{edge cloud} \cite{MEC_white_paper}. Owing to the close distance from the mobile device to the cloud server, MEC has the potential to fulfill the critical end-to-end delay requirement of 5G networks. Moreover, through mobile edge computation offloading (MECO), the energy consumption of mobile devices can be also reduced by offloading intensive computation workload to the proximate MEC server for execution \cite{MEC_survey}.

The minimization of end-to-end delay and energy consumption in the MECO technique requires joint allocation of communication and computation resources among mobile devices and MEC servers. Recent years have seen lots of studies on this topic for both single-user \cite{Single_user_1,Single_user_2,Single_user_3,Single_user_4,Single_user_5,MCC_energy,MEC_Partial_computation_offloading,Energy_MCC_stochastic_channel} and multi-user \cite{Multi_user_1,Multi_user_2,Multi_user_3,Multi_user_4,Multi_user_5, Multi_cell} MECO systems. In \cite{Single_user_1} and \cite{Single_user_2}, the authors have derived the optimal resource allocation solution for a single-user MECO system with multiple elastic tasks to minimize the average execution latency of all tasks under the transmit power constraint. On the other hand, to reduce the total energy consumption under a given latency requirement, the authors in \cite{Single_user_3} have derived the optimal threshold-based offloading policy with joint communication and computation resource allocation and the authors in \cite{Single_user_4} have proposed the optimal mode selection between local computing and cloud computing. Furthermore, a delay-optimal problem in a single-user MECO system with a prescribed resource utilization constraint has been studied in \cite{Single_user_5}, and a polynomial-time approximate solution with guaranteed performance has been developed herein. Optimal resource allocation and offloading decision policy has been further investigated to minimize the weighted-sum mobile energy consumption under the computation latency constraint \cite{Multi_user_1}. Besides, an online joint communication and computation resource management algorithm for a multi-user MECO system has been developed to minimize the long-term average weighted-sum energy consumption of mobile devices and the cloud server under the buffer stability constraint \cite{Multi_user_2}. A stochastic task arrival model based on the Lyapunov optimization algorithm has been proposed to solve the energy-latency tradeoff problem for a multi-user MECO system \cite{Multi_user_3}. In addition, to minimize the total energy consumption and offloading latency, game-theoretic techniques have been applied to develop the distributed algorithm, which is able to achieve a Nash equilibrium \cite{Multi_user_4}. Moreover, to cope with the bursty task arrivals, the MEC server can be integrated with uplink-downlink transmission scheduling to minimize the average latency \cite{Multi_user_5}.

Most aforementioned works on multi-user MECO systems focus on the binary computation offloading strategy, i.e., the computation task is executed either at the mobile device or at the edge cloud. Although the pioneering work in \cite{Multi_user_1} has studied the energy-efficient partial computation offloading, the latency-minimization issue is not discussed therein, which is a more urgent design target for 5G networks. Inspired by this, we investigate the latency-minimization problem in a multi-user MECO system with partial computation offloading in this paper. We assume that mobile devices have a large volume of raw data that are required to be compressed and uploaded to the edge cloud for analysis and storage. This considered scenario is corresponding to the surveillance and security application where massive online monitoring data should be timely transmitted to and analyzed by a central unit. Our design objective is to minimize the weighted-sum delay of all devices under the limited communication and computation resource constraints. According to where the data is compressed, we propose three different models: $\emph{local compression}$ where data is compressed only at mobile devices, $\emph{edge cloud compression}$ where data is transmitted to the edge cloud for compression, and $\emph{partial compression offloading}$ where partial data is compressed locally while the other part is compressed at the edge cloud. The main contributions of this work are summarized as follows.
\begin{itemize}
	\item For the local compression model, we formulate a convex optimization problem to minimize the weighted-sum delay of all devices under the communication resource constraint. Both closed-form
          expressions of optimal resource allocation and minimum weighted-sum delay are derived, and some inherent insights are also highlighted.
    \item For the edge cloud compression model, we analyze the task completion process by modeling a joint resource allocation problem with the constraints of both communication and computation resources.
          Then the closed-form solution and the minimum weighted-sum delay of all devices can be obtained by utilizing the Lagrange multiplier method.
    \item For the partial compression offloading model, we first formulate a piecewise optimization problem and then derive the optimal data segmentation strategy in a piecewise structure. Based on this
          result, we transform the original problem into a piecewise convex problem and develop an optimal resource allocation solution based on the sub-gradient algorithm.
    \item To yield more insights into the partial compression offloading model, we investigate a common scenario where
          communication resource is adequate while computation resource is limited. In this specific case, the delay expression of each device can be simplified and the closed-form solution of the piecewise optimization problem can be derived. It is also verified by numerical simulation that the proposed solution for this specific scenario can achieve a near-optimal performance in general scenarios.
\end{itemize}

The rest of this paper is organized as follows. In Section \uppercase\expandafter{\romannumeral2}, we introduce the multi-user MECO system and the three different compression models. In Section \uppercase\expandafter{\romannumeral3}, we investigate the local compression model and present a closed-form solution to the latency-minimization problem. The resource allocation problem for the edge cloud compression model is analyzed in Section \uppercase\expandafter{\romannumeral4}. Section \uppercase\expandafter{\romannumeral5} investigates the partial compression offloading model and the closed-form solution for a specific scenario is also analyzed in this section. Simulation results are presented in Section \uppercase\expandafter{\romannumeral6} and the whole paper is concluded in Section \uppercase\expandafter{\romannumeral7}.

\section{System Model}
In this section, we first introduce the multi-user MECO system. After that, we analyze the end-to-end delay of each device in the three models of local compression, edge cloud compression, and partial compression offloading, respectively.

\subsection{Multi-user MECO System}
\begin{figure}[!htp]
	\centering
	\includegraphics[width=0.8\textwidth]{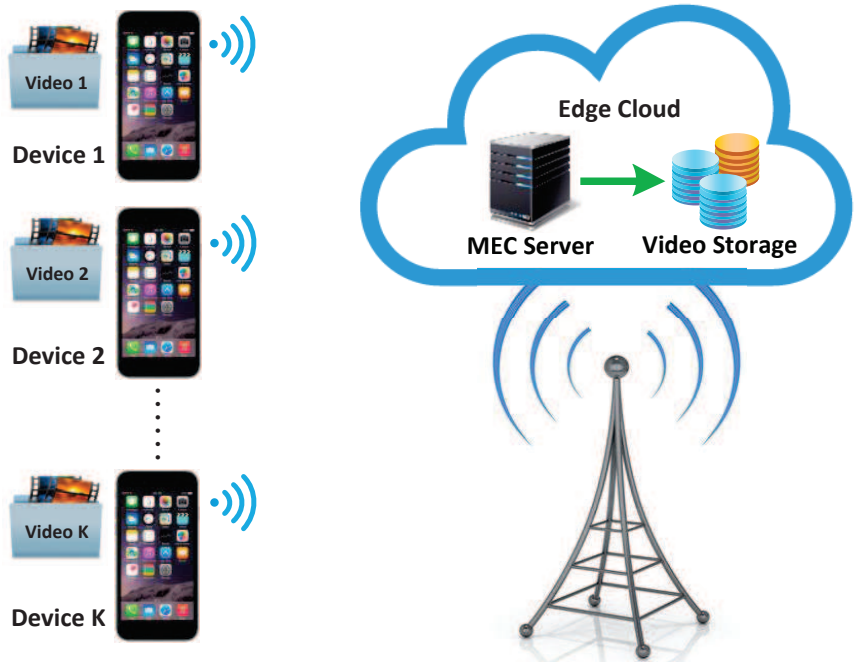}
	\caption{Multi-user MECO system model.}	
	\label{System model}
\end{figure}

As depicted in Fig. \ref{System model}, we consider a multi-user MECO system consisting of one edge cloud platform and $K$ single-antenna mobile devices, denoted by a set $\mathcal{K}=\{1, 2, \cdots, K \}$. The edge cloud can be regarded as a data center that is connected by mobile devices through wireless channels. Each device has a volume of data, such as a raw video, that needs to be compressed and stored in the edge cloud.\footnote{In this paper, we take the video compression as the example for the following analysis whereas our proposed framework can by extended into any data analytical system.} The mobile device $k$ in the system can be characterized by two parameters, i.e., the size of raw video $L_k>0$ (in bits) and the CPU compression capacity $V_k^{\text{d}}$ (in bits/s). Denote the total compression capacity of the edge cloud as $V^{\text{c}}$ (in bits/s), which can be allocated to all devices. That is, device $k$ will be allocated $V_k^{\text{c}}$ computational resource with constraint $\sum_{k=1}^{K} V_k^{\text{c}} \le V^{\text{c}}$.  Additionally, some other reasonable assumptions used in this paper are described as follows.
\begin{itemize}
    \item The edge cloud has perfect knowledge of the channel gains and the size of videos of all devices, which is required by the centralized scheduler.
	\item To guarantee that all videos can be compressed at the edge cloud simultaneously, we require that all devices and the edge cloud utilize the same video compression technology, such as MPEG4.
          The compression ratio is denoted as $\beta\in\left(0,1\right)$, i.e., one-bit raw video data will be compressed into $\beta$ bit.
	\item The delay for video segmenting, stitching, and storing can be reasonably neglected since they are much shorter than both communication and computational delays.
\end{itemize}

\subsection{Multiple-Access Model}
We apply a time division multiple access (TDMA) method for the channel access. In this method, one time frame is divided into $K$ time slots, which will be allocated to $K$ devices. For convenience, we normalize the duration of time-slot allocated to device $k$ as $t_k$ $\left(t_k \in [0, 1]\right)$. Note that the length of each time frame is short enough (e.g., 10 ms in LTE standards), which can be reasonably ignored when calculating the end-to-end delay of each device.

Let $h_k(i)$ denote the channel gain of device $k$ in time-slot $i$, which is a random variable and independently and identically distributed (i.i.d.) across the time-slot. Define $p_k$ as the transmission power of device $k$. Then the achievable data rate (in bits/s) in time-slot $i$ can be expressed as
\begin{equation}
    r_k(i) =  B \log_2 \left(1 + \frac{p_k \left|h_k(i)\right|^2}{N_0}\right), \label{1}
\end{equation}
where $B$ and $N_0$ are the bandwidth and the variance of additive white Gaussian noise (AWGN), respectively.

\subsection{Local Compression Model}
In the local compression model, each raw video is compressed locally and then transmitted to the edge cloud for storage. There are two kinds of delay in this model:
\begin{itemize}
	\item The delay for compressing the raw video $L_k$ bits at device $k$, $D_{\text{comp},k}^{\text{d}} = \dfrac{L_k}{V_k^{\text{d}}}$.
	\item The delay for transmitting the compressed video $\beta L_k$ bits to the edge cloud, $D_{\text{tran},k}^{\text{d}}$.
\end{itemize}

Assume that it takes at least $N_k^*$ time slots for device $k$ to transmit the compressed video $k$ of $\beta L_k$ bits to the edge cloud, where $N_k^*$ satisfies
\begin{equation}
    N_k^*= \arg \min \left\{ N : \sum_{i=1}^N r_k(i) \ge \frac{\beta L_k} {T_k}\right\},\label{average slot}
\end{equation}
where $T_k$ is the duration of time-slot allocated to device $k$. Note that the transmission rate $r_k(i)~(i=1,2,\cdots,N)$, as a function of the random channel gain $h_k(i)$, is also a random variable and i.i.d. across time slots, therefore $N_k^*$ is also a random variable. However, according to \cite {Rate_reference} and the martingale theory \cite{Martingale}, we can evaluate the average transmission delay as
\begin{equation}
D_{\text{tran},k}^{\text{d}} = \mathbb{E}_{ {\boldsymbol{h}}} \left\{ T_k N_k^* \right\} = \dfrac{\beta L_k} { \mathbb{E}_{ {\boldsymbol{h}}} \left\{ t_k r_k\right\}}=\dfrac{\beta L_k} { t_k\mathbb{E}_{ {\boldsymbol{h}}}\left\{ r_k\right\}}, \label{2}
\end{equation}
where $\mathbb{E}_{{\boldsymbol{h}}} \left\{ \cdot \right\}$ is the expectation over the channel gain $h_k(i)$.

For ease of notation, we define $R_k=\mathbb{E}_{ {\boldsymbol{h}}}\left\{ r_k\right\}$, which can be regarded as the average data rate of device $k$ across time slots. Moreover, we assume that the transmission can happen only after all data of the raw video is completely compressed. Then the end-to-end delay for device $k$ to complete its task can be expressed as
\begin{equation}
    D_k = D_{\text{comp},k}^{\text{d}} + D_{\text{tran},k}^{\text{d}} = \frac{L_k}{V_k^{\text{d}}} + \frac{\beta L_k}{t_k R_k}. \label{3}
\end{equation}

\subsection{Edge Cloud Compression Model}
In the edge cloud compression model, each device directly uploads its raw video to the edge cloud without any compression. Thereafter, the edge cloud compresses all raw videos in parallel by optimally allocating its computation resource. Similar to the local compression model, there also exist two kinds of delay in this model:
\begin{itemize}
	\item The delay for transmitting the raw video $L_k$ bits to the edge cloud, $D_{\text{tran},k}^{\text{c}} = \dfrac{L_k}{t_k R_k}$.
	\item The delay for compressing the raw video $L_k$ bits at the edge cloud, $D_{\text{comp},k}^{\text{c}} = \dfrac{L_k}{V_k^{\text{c}}}$.
\end{itemize}

Correspondingly, we require that the edge cloud can start compressing a raw video only after completely receiving its whole data. Then the end-to-end delay for device $k$ to complete its task can be written as
\begin{equation}
    D_k = D_{\text{tran},k}^{\text{c}} + D_{\text{comp},k}^{\text{c}} = \frac{L_k}{t_k R_k}+\frac{L_k}{V_k^{\text{c}}}. \label{4}
\end{equation}

\subsection{Partial Compression Offloading Model}
\begin{figure}[!htp]
	\centering
	\includegraphics[width=0.8\textwidth]{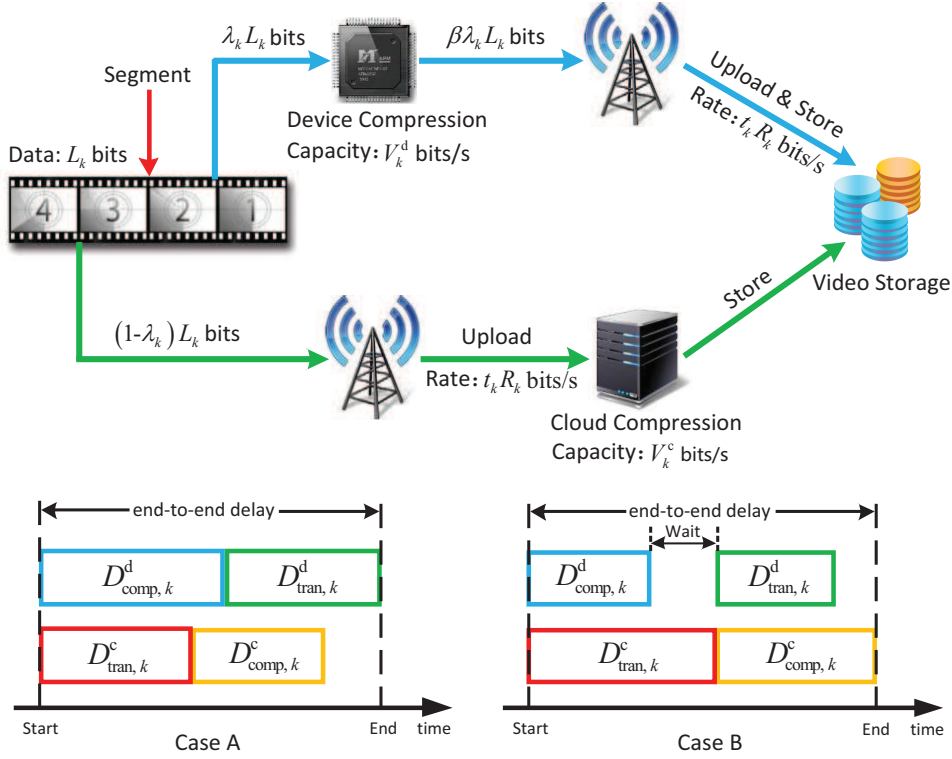}
	\caption{The whole process of partial compression offloading.}	
	\label{Partial compression offloading}
\end{figure}

In the local compression model, the local compression delay, $D_{\text{comp},k}^{\text{d}}$, would be dominant if the speed of device CPU is limited, i.e., corresponding to the wireless video monitoring camera. On the other hand, in the edge cloud compression model, the transmission delay, $D_{\text{tran},k}^{\text{c}}$, would be dominant if the channel bandwidth is limited. Obviously, both two models are not optimal in terms of end-to-end delay minimization if the video can be partially compressed at the mobile device and partially compressed at the edge cloud. Motivated by this, in this subsection, we propose a partial compression offloading model, in which each raw video can be partitioned into two parts with one compressed locally while the other offloaded for edge compression. Let us denote the proportion of video $k$ that is compressed at the mobile device as $\lambda_k\in \left[0,1\right]$. Then, we can introduce the detailed procedure of partial compression offloading in three steps, as depicted in Fig. \ref{Partial compression offloading}.
\begin{itemize}
	\item Mobile device $k$ compresses $\lambda_k L_k$ bits of the raw video locally and then transmits the compressed data of $\beta \lambda_k L_k$ bits to the edge cloud.
	\item Mobile device $k$ transmits the remaining $(1-\lambda_k)L_k$ bits to the edge cloud. Then the edge cloud compresses this part of raw video itself.
    \item Finally, the edge cloud combines two parts of compressed video and stories into the data-storage center.
\end{itemize}

In the above three steps, there exist four kinds of delay as follows.
\begin{itemize}
	\item The delay for compressing $\lambda_k L_k$ bits at mobile device $k$: $D_{\text{comp},k}^{\text{d}} = \dfrac{\lambda_k L_k}{V_k^{\text{d}}}$.
	\item The delay for transmitting the local compressed part: $D_{\text{tran},k}^{\text{d}} = \dfrac{\beta \lambda_k L_k}{t_k R_k}$.
	\item The delay for transmitting the uncompressed part: $D_{\text{tran},k}^{\text{c}} = \dfrac{(1-\lambda_k)L_k}{t_k R_k}$.
	\item The delay for compressing $(1-\lambda_k)L_k$ bits at the edge cloud: $D_{\text{comp},k}^{\text{c}} = \dfrac{(1-\lambda_k)L_k}{V_k^{\text{c}}}$.
\end{itemize}

Since each device has only one channel for data transmission, either local compression part or edge cloud compression part can be transmitted at any moment while not simultaneously. Therefore, as depicted in Fig. \ref{Partial compression offloading}, two cases will happen. The first one corresponds to $D_{\text{comp},k}^{\text{d}} \ge D_{\text{tran},k}^{\text{c}}$, where the transmission for local compressed video can start immediately at the end of the local compression. The second one corresponds to $D_{\text{comp},k}^{\text{d}} < D_{\text{tran},k}^{\text{c}}$, where the transmission for local compressed video must wait until the transmission for the edge cloud compression part ends. Therefore, the end-to-end delay of device $k$ in this model can be written as
\begin{equation}
	D_k = \left \{
    \begin{aligned}
        &\max \left \{ D_{\text{comp},k}^{\text{d}} + D_{\text{tran},k}^{\text{d}},D_{\text{tran},k}^{\text{c}} + D_{\text{comp},k}^{\text{c}}
              \right\}, &\text{if}~D_{\text{comp},k}^{\text{d}} \ge D_{\text{tran},k}^{\text{c}},\\
        &D_{\text{tran},k}^{\text{c}} + \max \left \{ D_{\text{tran},k}^{\text{d}},D_{\text{comp},k}^{\text{c}} \right \}, &\text{if}~D_{\text{comp},k}^{\text{d}} < D_{\text{tran},k}^{\text{c}}.\label{5}
	\end{aligned} \right.
\end{equation}

In the next three sections, we will develop optimal joint communication and computation resource allocation algorithms to minimize the weighted-sum delay of all devices for the three different models, respectively.

\section{Optimal Solution to the Local Compression Model}
In this section, we first formulate the latency-minimization problem for the local compression model and then derive the closed-form expressions for both optimal solution and minimum weighted-sum delay of all devices.

\subsection{Problem Formulation}
We aim at minimizing the weighted-sum delay of all devices, $\sum_{k=1}^K \alpha_k D_k$, where the positive weight factors $\{\alpha_k\}$ account for the fairness among devices and satisfy $\sum_{k=1}^K \alpha_k = 1$. Based on the end-to-end delay expression in (\ref{3}), we have the following optimization problem for the local compression model.
\begin{prob}$\emph{(Local~Compression)}$
\begin{subequations}
	\begin{eqnarray}
	&\min\limits_{\{t_k\}}&\sum_{k=1}^K \alpha_k\left(\frac{L_k}{V_k^{\text{d}}}+\frac{\beta L_k}{t_k R_k}\right),\label{eq:P1.a}\\
	&\text{s.t.}&\sum_{k=1}^K t_k \le 1,~t_k \ge 0, \label{eq:P1.b}
	\end{eqnarray}
\end{subequations}
\end{prob}
where (\ref{eq:P1.b}) is the overall communication resource constraint of all devices.

\subsection{Optimal Solution}
It can be easily verified that Problem 1 is convex and the Slater's condition can be satisfied, implying that strong duality holds. Thus, Problem 1 can be solved by the Karush-Kuhn-Tucker (KKT) conditions. The Lagrange function can be expressed as
\begin{equation}
    L_{\text{L}} = \sum_{k=1}^K \alpha_k \left(\frac{L_k}{V_k^{\text{d}}} + \frac{\beta L_k}{t_k R_k}\right) + \nu \left( \sum_{k=1}^K t_k - 1 \right), \label{7}
\end{equation}
where $\nu \ge 0$ is the Lagrange multiplier associated with the constraint (\ref{eq:P1.b}). Let $\left\{t_k^{*(1)}\right\}$ denote the optimal solution for Problem 1. Then applying KKT conditions leads to the following necessary and sufficient conditions
\begin{equation}
    \frac{\partial L_{\text{L}}}{\partial t_k^{*(1)}} = -\frac{\alpha_k \beta L_k}{R_k \left( t_k^{*(1)} \right)^2 }+\nu^*
    \left\{ \begin{aligned}
            &>0,~~t_k^{*(1)}=0,\\
            &=0,~~t_k^{*(1)}>0,
            \end{aligned}
    \right. \label{8}
\end{equation}
\begin{equation}
    \nu^*\left(\sum_{k=1}^K t_k^{*(1)} - 1\right)=0,~\sum_{k=1}^K t_k^{*(1)} \le 1,~\nu^*\ge 0. \label{9}
\end{equation}
Based on these conditions, we can derive the following optimal solution.
\begin{thm}
The optimal solution solving Problem 1 of the local compression model is given by
\begin{equation}
     t_k^{*(1)}=\dfrac{\sqrt{\dfrac{\alpha_k L_k}{R_k}}}{\sum_{k=1}^K{\sqrt{\dfrac{\alpha_k L_k}{R_k}}}},~\forall k \in \mathcal{K}. \label{10}
\end{equation}
\end{thm}
\begin{rem}
Theorem 1 reveals that the optimal time-slot allocated to device $k$ is determined by the corresponding weight factor, size of raw video, and channel capacity. The weight factor, $\alpha_k$, can be interpreted as the level of importance for device $k$. The larger the value of $\alpha_k$ is, the more the time-slot should be allocated to device $k$ to minimize the whole system delay. Furthermore, more time slot should be allocated to device $k$ if the video size, $L_k$, becomes larger or the channel capacity, $R_k$, gets smaller.
\end{rem}

Based on the above solution, we can derive the minimum system delay (i.e., weighted-sum delay of all devices) in a closed-form way, as
\begin{equation}
	\begin{aligned}
	D_{\text{sys}}^{\text{L}}
        &= \sum_{k=1}^K{ \frac{\alpha_k L_k}{V_k^{\text{d}}} + \sum_{i=1}^K{\left(\sqrt{\alpha_i\frac{\beta L_i}{R_i}}\sum_{j=1}^K{\sqrt{\alpha_j\frac{\beta L_j}{R_j}}}\right)}}\\
        &= \sum_{k=1}^K{ \frac{\alpha_k L_k}{V_k^{\text{d}}}} + \beta \sum_{i=1}^K{\sum_{j=1}^K{\sqrt{\frac{\alpha_i \alpha_j L_i L_j}{R_iR_j}}}.}
	\end{aligned} \label{11}
\end{equation}

\section{Optimal Solution to the Edge Cloud Compression Model}
In this section, we analyze the latency-minimization problem for the edge cloud compression model and devise the joint optimal communication and computation resource allocation algorithm.
\subsection{Problem Formulation}
To minimize the weighted-sum delay of all devices, we have the following problem for the edge cloud compression model.
\begin{prob}$\emph{(Edge~Cloud~Compression)}$
    \begin{subequations}
    \begin{eqnarray}
    &\min\limits_{\{t_k, V_k^{\text{c}}\}}&\sum_{k=1}^K \alpha_k \left(\frac{L_k}{t_kR_k}+\frac{L_k}{V_k^{\text{c}}}\right),\label{eq:P2.a}\\
    &{\text{s.t.}}&\sum_{k=1}^K t_k \le 1,~t_k \ge 0, \label{eq:P2.b}\\
    &&\sum_{k=1}^K V_k^{\text{c}} \le V^{\text{c}},~V_k^{\text{c}} \ge 0, \label{eq:P2.c}
    \end{eqnarray}
    \end{subequations}
\end{prob}
where constraints ({\ref{eq:P2.b}}) and ({\ref{eq:P2.c}}) imply that the overall communication and computation resources allocated to mobile devices cannot exceed the corresponding limitations.

\subsection{Optimal Solution}
Fortunately, Problem 2 is also convex since each component in (\ref{eq:P2.a}) is convex on $t_k$ and $V_k^{\text{c}}$. Therefore, Problem 2 can be optimally solved using the KKT conditions. The Lagrange function can be written as
\begin{equation}
     L_{\text{E}}=\sum_{k=1}^K \alpha_k \left(\frac{L_k}{t_kR_k}+\frac{L_k}{V_k^{\text{c}}}\right) + \xi\left(\sum_{k=1}^K t_k -1\right)+\chi \left(\sum_{k=1}^K V_k^{\text{c}}-V^{\text{c}}\right),\label{13}
\end{equation}
where $\xi \ge 0$ and $\chi \ge 0$ are the Lagrange multipliers associated with constraints (\ref{eq:P2.b}) and (\ref{eq:P2.c}), respectively. Let $\left\{t_k^{*(2)},V_k^{\text{c}*(2)}\right\}$ denote the optimal solution for Problem 2. Then the necessary and sufficient conditions based on the KKT conditions can be expressed as
\begin{equation}
    \frac{\partial L_{\text{E}}}{\partial t_k^{*(2)}}=-\frac{\alpha_k L_k}{ R_k\left(t_k^{*(2)}\right)^2}+\xi^*
    \left\{
    \begin{aligned}
        &>0,~~t_k^{*(2)}=0,\\
        &=0,~~t_k^{*(2)}>0,
    \end{aligned}
    \right. \label{14}
\end{equation}
\begin{equation}
    \frac{\partial L_{\text{E}}}{\partial V_k^{\text{c}*(2)}}=-\frac{\alpha_k L_k}{\left(V_k^{\text{c}*(2)}\right)^2}+\chi^*
    \left\{
    \begin{aligned}
        &>0,~~V_k^{\text{c}*(2)}=0,\\
        &=0,~~V_k^{\text{c}*(2)}>0,
    \end{aligned}
    \right. \label{15}
\end{equation}
\begin{equation}
    \xi^*\left(\sum_{k=1}^K t_k^{*(2)} - 1\right)=0,~\sum_{k=1}^K t_k^{*(2)}\le 1,~\xi^* \ge 0,\label{16}
\end{equation}
\begin{equation}
    \chi^*\left(\sum_{k=1}^K V_k^{\text{c}*(2)}- V^{\text{c}}\right)=0,~\sum_{k=1}^K V_k^{\text{c}*(2)} \le V^{\text{c}},~\chi^* \ge 0. \label{17}
\end{equation}
By solving the above equations, we can obtain the optimal solution for Problem 2, as shown in Theorem 2.
\begin{thm}
The optimal solution for Problem 2 of the edge cloud compression model is given by
\begin{equation}
	\left\{
    \begin{aligned}
        &t_k^{*(2)} = \dfrac{\sqrt{\dfrac{\alpha_k L_k}{R_k}}}{\sum_{k=1}^K{\sqrt{\dfrac{\alpha_k L_k}{R_k}}}},~\forall k \in \mathcal{K},\\
        &V_k^{\text{c}*(2)} = \dfrac{\sqrt{\alpha_k L_k}}{\sum_{k=1}^K{\sqrt{\alpha_k L_k}}}V^{\text{c}},~\forall k \in \mathcal{K}.
	\end{aligned}
    \right. \label{19}
\end{equation}
\end{thm}
\begin{rem}
From Theorem 2, we can see that the optimal time-slot allocated to each device in the edge cloud compression model has the same expression as that in the local compression model, and the optimal cloud compression capacity allocated to each device is determined by the corresponding weight factor and video size. Similarly, the weight factor $\alpha_k$ is positively related to the allocated resources since it reflects the level of importance for device $k$. Specially, in case that each device has the same weight, i.e., $\alpha_k=\dfrac{1}{K}, \forall k \in \mathcal{K}$, the bigger the video size is, the more the time-slot and edge cloud compression capacity should be allocated to the device for achieving the minimum system delay. Again, we can express the minimum system delay in a closed-form way, as
\begin{equation}
	\begin{aligned}
	    D_{\text{sys}}^{\text{E}}
        &= \sum_{i=1}^K{\left( \sqrt{\alpha_i \frac{L_i}{R_i}} \sum_{j=1}^K{ \sqrt{\alpha_j \frac{L_j}{R_j}}} \right)} +
           \sum_{i=1}^K{\left(\sqrt{\alpha_i L_i}\sum_{j=1}^K{\frac{\sqrt{\alpha_j L_j}}{V^{\text{c}}}} \right)}\\
        &= \sum_{i=1}^K{\sum_{j=1}^K{\sqrt{\alpha_i \alpha_j L_i L_j}\left(\sqrt{\frac{1}{R_i R_j}}+\frac{1}{V^{\text{c}}}\right)}}.
	\end{aligned} \label{20}
\end{equation}
\end{rem}

\section{Optimal Solution to the Partial Compression Offloading Model}
In the above sections, we have analyzed the optimal communication and computation resource allocations for the local compression model and the edge cloud compression model, respectively. In this section, we shall investigate the latency-minimization problem for the partial compression offloading model. The problem studied in this section is more generic in that it fully utilizes the computation resource in both mobile devices and the edge cloud, which can further reduce the system delay. In the following, we shall first formulate the latency-minimization problem and then derive the optimal video segmentation strategy in a piecewise structure. After that, we will transform the original problem into a piecewise convex problem and develop a sub-gradient algorithm to find the optimal solution efficiently. Finally, the closed-form solution in a specific scenario will be also devised.

\subsection{Problem Formulation}
In the partial compression offloading model, each raw video could be partially compressed at the mobile device and partially compressed at the edge cloud. Therefore, both communication and computation resources should be jointly allocated and the optimization problem can be formulated as
\begin{prob}$\emph{(Partial~Compression~Offloading)}$
    \begin{subequations}
    \begin{eqnarray}
    &\min\limits_{\{t_k,V_k^{\text{c}},\lambda_k\}}&\sum_{k=1}^K \alpha_k D_k,\label{eq:P3.a}\\
    &{\text{s.t.}}&\sum_{k=1}^K t_k \le 1,~t_k \ge 0, \label{eq:P3.b}\\
    &&\sum_{k=1}^K V_k^{\text{c}} \le V^{\text{c}},~V_k^{\text{c}} \ge 0,\label{eq:P3.c}\\
    &&0 \le \lambda_k\le 1,~\forall k \in \mathcal{K}. \label{eq:P3.d}
    \end{eqnarray}
    \end{subequations}
\end{prob}
Notice that $D_k$ is a piecewise function given in (\ref{5}), which can be rewritten in a more detailed way, as
\begin{equation}
	D_k =
    \left\{
    \begin{aligned}
        &\max \left \{ \frac{\lambda_k L_k}{V_k^{\text{d}}} + \frac{\beta \lambda_k L_k}{t_k R_k},\frac{(1-\lambda_k)L_k}{t_kR_k} + \frac{(1-\lambda_k)L_k}{V_k^{\text{c}}} \right\},
            &\text{if}~\lambda_k \ge \frac{V_k^{\text{d}}}{V_k^{\text{d}} + t_k R_k},\\
        &\frac{(1-\lambda_k)L_k}{t_k R_k}+\max \left\{\frac{\beta \lambda_k L_k}{t_k R_k},\frac{(1-\lambda_k)L_k}{V_k^{\text{c}}} \right\} ,
            &\text{if}~\lambda_k<\frac{V_k^{\text{d}}}{V_k^{\text{d}}+t_k R_k}.\\
	\end{aligned}
    \right. \label{21}
\end{equation}

\subsection{Optimal Segmentation Strategy and Problem Transformation}
It can be seen that the $D_k$ expression in (\ref{21}) is complicated with 3$K$ variables such that Problem 3 is hard to be solved directly. In the following, we will determine the optimal $\lambda_k^{*}$ while keeping $t_k$ and $V_k^{\text{c}}$ fixed. First, let us define a geometric mean $\sqrt{\beta V_k^{\text{d}}V_k^{\text{c}}}$, which is referred to as the \emph{average compression capacity} for compressing video $k$. Correspondingly, $t_k R_k$ can be regarded as the \emph{average communication capacity} for transmitting video $k$. Then the optimal video segmentation strategy has the following piecewise structure, as presented in Lemma \ref{Lemma1}.
\begin{lem}
 Given the sets of $\{t_k\}$ and $\left\{V_k^{\text{c}}\right\}$, the optimal video segmentation strategy for each device is given by
	\begin{equation}
	\lambda_k^* =
    \left\{
	\begin{aligned}
	   &\frac{V_k^{\text{d}}\left(t_kR_k+V_k^{\text{c}}\right)}{V_k^{\text{d}}V_k^{\text{c}}(1+\beta)+t_kR_k(V_k^{\text{d}}+V_k^{\text{c}})},
            &\text{if}~t_k R_k \ge \sqrt{\beta V_k^{\text{d}} V_k^{\text{c}}},\\
	   &\frac{V_k^{\text{d}}}{V_k^{\text{d}}+t_kR_k},
            &\text{if}~t_k R_k < \sqrt{\beta V_k^{\text{d}} V_k^{\text{c}}}.\label{22}
	\end{aligned}
	\right.
	\end{equation} \label{Lemma1}
\end{lem}
\begin{IEEEproof}
	Please refer to Appendix A.
	\end{IEEEproof}
\begin{rem}
 The optimal video segmentation strategy shown in Lemma 1 is determined by comparing the \emph{average communication capacity} with the \emph{average compression capacity} of each device. In the case that the \emph{average communication capacity} dominates the \emph{average compression capacity}, i.e., $t_k R_k \ge \sqrt{\beta V_k^{\text{d}} V_k^{\text{c}}}$, the computation resource is the bottleneck of delay minimization for device $k$, and therefore we should make full use of the computation resource. In this case, $\lambda_k^*$ satisfies $D_{\text{comp},k}^{\text{d}} \ge D_{\text{tran},k}^{\text{c}}$ while $D_{\text{comp},k}^{\text{d}} + D_{\text{tran},k}^{\text{d}} = D_{\text{tran},k}^{\text{c}} + D_{\text{comp},k}^{\text{c}}$. On the contrary, in the case that $t_k R_k < \sqrt{\beta V_k^{\text{d}} V_k^{\text{c}}}$,  the communication resource is the main bottleneck of delay minimization. Therefore, we need to fully utilize the communication resource to minimize the end-to-end delay of each device. In this case, $\lambda_k^*$ fulfills that $D_{\text{comp},k}^{\text{d}} = D_{\text{tran},k}^{\text{c}}$ while $D_{\text{comp},k}^{\text{c}} < D_{\text{tran},k}^{\text{d}}$.
\end{rem}

By substituting the optimal video segmentation strategy into (\ref{21}), the end-to-end delay of device $k$ can be written as
\begin{equation}
   \widehat{D}_k=
   \left\{
   \begin{aligned}
        &\frac{L_k}{t_k R_k} \frac{\left(t_kR_k+V_k^{\text{c}}\right) \left(t_k R_k+\beta V_k^{\text{d}}\right)}{V_k^{\text{d}}V_k^{\text{c}}(1+\beta)+t_k R_k(V_k^{\text{d}}+V_k^{\text{c}})}
            \triangleq \widehat{D}_{k,1}, & &\text{if}~t_k R_k \ge \sqrt{\beta V_k^{\text{d}} V_k^{\text{c}}}, \\
        &\frac{L_k}{t_k R_k}\frac{t_k R_k+\beta V_k^{\text{d}}}{V_k^{\text{d}}+t_k R_k} \triangleq \widehat{D}_{k,2},
                                          & &\text{if}~t_k R_k < \sqrt{\beta V_k^{\text{d}} V_k^{\text{c}}}. \\
   \end{aligned}
   \right. \label{23}
\end{equation}
Then Problem 3 can be equivalently converted to the following problem.
\begin{prob}$\emph{(Equivalent Problem of Problem 3).}$
    \begin{subequations}
    \begin{eqnarray}
    &\min\limits_{\{t_k,V_k^{\text{c}}\}}&\sum_{k=1}^K \alpha_k \widehat{D}_k, \label{eq:P4}\\
    &\text{s.t.}&\text{(\ref{eq:P3.b})}~\text{and}~\text{(\ref{eq:P3.c})}.
    \end{eqnarray}
    \end{subequations}
\end{prob}
\begin{thm}
    Problem 4 is a piecewise convex optimization problem.
\end{thm}
\begin{IEEEproof}
	Please refer to Appendix B.
\end{IEEEproof}

\subsection{Optimal Resource Allocation Algorithm}
The key challenge of Problem 4 is that the $\widehat{D}_k$ expression in (\ref{23}) is continuous but non-differential (or non-smooth) at $t_k R_k = \sqrt{\beta V_k^{\text{d}} V_k^{\text{c}}}$. Moreover, the partial derivatives of both $\widehat{D}_{k,1}$ and $\widehat{D}_{k,2}$ on $t_k$ have quartic forms. Therefore, classical KKT conditions cannot be directly applied to solve this problem and it is rather difficult to find its closed-form solution. In the following, we will develop an effective algorithm to optimally solve it, which is based on the sub-gradient method for common non-differential convex problems \cite{Convex_Optimization_II}. For ease of notation, we shall first define the following auxiliary variables.
\begin{itemize}
	\item Define a vector of independent resource variables, as ${\bm{x}} = \left[t_1,t_2,\cdots,t_K,V^{\text{c}}_1,V^{\text{c}}_2,\cdots,V^{\text{c}}_K\right]$.
	\item Define the weighted-sum delay of all devices, as $F = \sum_{k=1}^{K}\alpha_k\widehat{D}_k$.
\end{itemize}
After that, we denote the sub-gradient function of $\widehat{D}_k$ as $\partial \widehat{D}_k=\left[\dfrac{\partial \widehat{D}_k}{\partial t_k},\dfrac{\partial \widehat{D}_k}{\partial V_k^{\text{c}}}  \right]$. Since it has been proved in Theorem 3 that $\widehat{D}_k$ can be written as $\max \left\{\widehat{D}_{k,1}, \widehat{D}_{k,2} \right\}$, the sub-gradient function  can be characterized as
\begin{equation}
    \dfrac{\partial \widehat{D}_k}{\partial t_k} \in
    \left\{
    \begin{aligned}
        & \frac{\partial \widehat{D}_{k,1}}{\partial t_k},
            & &\text{if}~t_k R_k > \sqrt{\beta V_k^{\text{d}} V_k^{\text{c}}}, \\
        & \left[\frac{\partial \widehat{D}_{k,2}}{\partial t_k}, \frac{\partial \widehat{D}_{k,1}}{\partial t_k}\right],
            & &\text{if}~t_k R_k = \sqrt{\beta V_k^{\text{d}} V_k^{\text{c}}}, \\
        & \frac{\partial \widehat{D}_{k,2}}{\partial t_k},
            & &\text{if}~t_k R_k < \sqrt{\beta V_k^{\text{d}} V_k^{\text{c}}},
	\end{aligned}
    \right. \label{25}
\end{equation}
\begin{equation}
    \dfrac{\partial \widehat{D}_k}{\partial V_k^{\text{c}}} \in
    \left\{
    \begin{aligned}
        & \frac{\partial \widehat{D}_{k,1}}{\partial V_k^{\text{c}}},
            & &\text{if}~t_k R_k > \sqrt{\beta V_k^{\text{d}} V_k^{\text{c}}}, \\
        & \left[\frac{\partial \widehat{D}_{k,1}}{\partial V_k^{\text{c}}}, \frac{\partial \widehat{D}_{k,2}}{\partial V_k^{\text{c}}}\right],
            & &\text{if}~t_k R_k = \sqrt{\beta V_k^{\text{d}} V_k^{\text{c}}}, \\
        & \frac{\partial \widehat{D}_{k,2}}{\partial V_k^{\text{c}}},
            & &\text{if}~t_k R_k < \sqrt{\beta V_k^{\text{d}} V_k^{\text{c}}}.
	\end{aligned}
    \right. \label{26}
\end{equation}

Based on the above analysis, we introduce the following theorem to solve Problem 4.
\begin{thm}
Problem 4 can be solved by the following iteration
\begin{equation}
	{\bm{x}}^{(n+1)} = {\bm{x}}^{(n)}-\phi_n {\bm{g}}^{(n)}, \label{27}
\end{equation}
\end{thm}		
where $\phi_n$ is the step size of the $n^{\text{th}}$ iteration and $\bm{g}$ is the sub-gradient function of $\sum_{k=1}^K \alpha_k \widehat{D}_k$, which is defined as
\begin{align}
	{\bm{g}} &=
    \left\{
	\begin{array}{ll}
		\partial\left(\sum_{k=1}^{K}\alpha_k \widehat{D}_k \right), &\text{subject to}~ \text{(\ref{eq:P3.b})}~\text{and}~\text{(\ref{eq:P3.c})},\\
		\partial\left(\sum_{k=1}^{K}t_k\right),&\text{if}~\sum_{k=1}^{K}t_k>1,\\
		\partial\left(\sum_{k=1}^{K}V_k^{\text{c}}\right),&\text{if}~\sum_{k=1}^{K}V_k^{\text{c}}>V^{\text{c}},
	\end{array}
	\right.\label{28}
\end{align}
where $\partial\left(\sum_{k=1}^{K}t_k\right)$ and $\partial\left(\sum_{k=1}^{K}V_k^{\text{c}}\right)$ are utilized as the obstacle functions.
\begin{IEEEproof}
	Please refer to Appendix C.
\end{IEEEproof}

Based on Theorem 4, we can efficiently solve Problem 4 by iteratively updating the communication and computation resource allocation, whose detailed procedures are presented in Algorithm 1.
    \begin{table}[!htp]
		\vspace{-2em}
		\begin{algorithm}[H]
			\caption{The sub-gradient algorithm for the partial compression offloading model}
			{\normalsize
				\begin{algorithmic}[1]
					
					\STATE \textbf{Initialize}
					
					\STATE ~~~~Initialize the maximum convergence tolerance $\epsilon > 0$.
					
					\STATE ~~~~Set the iteration index $n=0$.

                    \STATE ~~~~Set the initial resource allocation vector ${\bm{x}}^{(0)}$ that subjects to (\ref{eq:P3.b}) and (\ref{eq:P3.c}).
					
					\STATE ~~~~Calculate $F^{(0)} = \sum_{k=1}^K \alpha_k \widehat{D}_k^{(0)}$ and ${\bm{g}}^{(0)}$ according to (\ref{23}) and (\ref{28}).
					
					\STATE \textbf{Do}
					
					\STATE ~~~~Update the resource allocation vector by \\
					~~~~~~~~~~
					$	{\bm{x}}^{(n+1)} = {\bm{x}}^{(n)}-\phi_n {\bm{g}}^{(n)}$.
									
                    \STATE ~~~~Update $n=n+1$.

					\STATE ~~~~Calculate $F^{(n)} = \sum_{k=1}^K \alpha_k \widehat{D}_k^{(n)}$ and ${\bm{g}}^{(n)}$  according to (\ref{23}) and (\ref{28}).
					\STATE \textbf{Until} $|F^{(n)} - F^{(n-1)}| \le \varepsilon$.
			\end{algorithmic}}
		\end{algorithm}
	\end{table}
\vspace{-2em}

Now we discuss the convergence and the computational complexity of Algorithm 1. As we have proved in Appendix C, the vector $\bm{x}^{(n)}$ will linearly converge to the optimal solution $\bm{x}^*$ when $\epsilon \rightarrow 0$ \cite{Convex_Optimization_II}. On the other hand, the computational complexity of the sub-gradient algorithm mainly lies on the required number of iterations until convergence, which is determined by the maximum tolerance $\epsilon$. From \cite{Optimization_Methods_for_Large-Scale_Systems}, we can conclude that our proposed algorithm has a polynomial time complexity of $\mathcal{O}\left(\dfrac{1}{\epsilon^2}\right)$, which is desirable for  practical implementation.

\subsection{A Special Case}
The optimal solution developed in the above subsection is not in closed-form. To yield more insights into the partial compression offloading model, we further investigate a common scenario where the communication resource is adequate while the computation resource is limited, such as the typical sensor network or the machine-type communications. The key characteristic of this specific scenario is that the channel capacity is much greater than the device computation capacity, i.e., $R_k \gg V_k^{\text{d}}$. Moreover, most current mobile devices utilize the MEPG4 video compression technology whose compression ratio, $\beta$, is between $\dfrac{1}{50}$ and $\dfrac{1}{200}$ \cite{Video_compression}, resulting in the small size of local compressed video. Under these conditions, the delay for transmitting the local compressed video, $D_{\text{tran},k}^{\text{d}}$, can be neglected while comparing with the delay for compressing the local compression part of video, $D_{\text{comp},k}^{\text{d}}$. That is, $\dfrac{\beta \lambda_k L_k}{t_k R_k} \ll \dfrac{\lambda_k L_k}{V_k^{\text{d}}}$. Therefore, it is straightforward that the optimal video segmentation strategy in this case satisfies
\begin{equation}
    D_{\text{comp},k}^{\text{d}}=D_{\text{tran},k}^{\text{c}}+D_{\text{comp},k}^{\text{c}}. \label{29}
\end{equation}
Then applying the detailed delay expressions in Section \uppercase\expandafter{\romannumeral2}-E into (\ref{29}), we have the following optimal video segmentation strategy.
\begin{lem}
In the specific scenario of partial compression offloading, the optimal video segmentation strategy for each device is given by
\begin{equation}
	\overline{\lambda}_k^*=\frac{ V_k^{\text{d}} \left(t_kR_k+V_k^{\text{c}}\right)} { V_k^{\text{d}} V_k^{\text{c}} + t_k R_k \left(V_k^{\text{d}}+V_k^{\text{c}} \right)},~\forall k \in \mathcal{K}.\label{30}
\end{equation}
\end{lem}
Based on Lemma 2, the end-to-end delay of device $k$ can be written as
\begin{equation}
	\overline{D}_k=\frac{L_k\left(t_kR_k+V_k^{\text{c}}\right)}{V_k^{\text{d}}V_k^{\text{c}}+t_kR_k\left(V_k^{\text{d}}+V_k^{\text{c}} \right)}\label{31}.
\end{equation}
Substituting $\overline{D}_k$ into (\ref{eq:P4}), the convex Problem 4 can be solved by the KKT conditions. Therefore, the optimal solution for this specific scenario can be derived, as shown in Theorem 5.
\begin{thm}
The optimal solution for the specific scenario of partial compression offloading is given by
    \begin{equation}
    	\left\{
        \begin{aligned}
            &\overline{t}_k^* = \frac{\overline{V}_k^{\text{c}*}\left(\sqrt{\frac{\alpha_k L_k R_k}{\theta^*}}-V_k^{\text{d}} \right)^+}{R_k \left( V_k^{\text{d}} + \overline{V}_k^{\text{c}*} \right)},
                ~\forall k \in \mathcal{K},\\
            &\overline{V}_k^{\text{c}*} = \frac{\overline{t}_k^* R_k \left(\sqrt{\frac{\alpha_k L_k}{\omega^*}}-V_k^{\text{d}} \right)^+}{\overline{t}_k^*R_k + V_k^{\text{d}}},
                ~\forall k \in \mathcal{K},
    	\end{aligned}
        \right. \label{32}
    \end{equation}
where $(y)^+= \max\{y, 0\}$, $\theta^*$ and $\omega^*$ are the optimal value of Lagrange multipliers that satisfy the active communication and computation resource constraints $\sum_{k=1}^K \overline{t}_k^* = 1$ and $\sum_{k=1}^K \overline{V}_k^{\text{c}*} = V^{\text{c}}$, respectively.
\end{thm}
\begin{IEEEproof}
	Please refer to Appendix D.
\end{IEEEproof}
\begin{rem}
Theorem 5 reveals that the optimal time-slot and cloud compression capacity allocated to device $k$ is determined by the corresponding weight factor, size of raw video, channel capacity, and local compression capacity. As the video size $L_k$ increases, more communication and computation resources will be allocated to this device. However, less communication resource will be allocated if the communication capacity becomes larger. This result is consistent with the intuition that, to reduce the weighted-sum delay of all devices, the BS should allocate more communication resource to those devices with bad channels. Moreover, in case that the local compression capacity $V_k^{\text{d}}$ gets smaller, more communication and computation resources should be allocated to this device under the criterion of minimizing the system delay.
\end{rem}
\section{Numerical Results}
In this section, we will present numerical results to verify our analysis and validate the performance of the proposed algorithms. The simulation settings are as follows unless otherwise stated. The BS has a radius of 250 m. Each mobile device is randomly located in the system and can associate with the BS through one wireless channel. The weights for all devices are the same, i.e., $\alpha_k=\dfrac{1}{K}$ for all $k$ such that the system delay represents the average end-to-end delay of all devices. The channel gains between mobile devices and the edge cloud are generated according to i.i.d. Rayleigh random variables with unit variances. The transmission power is set equal for each device, i.e., $p_k=24$ dBm, $\forall k \in \mathcal{K}$.  The total bandwidth $B=$10 MHz. For each compression task, the video size and the  device compression capacity follow the uniform distribution with $L_k \in$ [10, 100] Mbits and $V_k^{\text{d}}\in$ [0.5, 2]  Mbps, respectively. All random variables are independent for different devices, modeling heterogeneous mobile compression capacity. The total compression capacity of the edge cloud $V^{\text{c}}$ is selected as 40 Mbps and the compression ratio $\beta$ is set as 0.01. Other major simulation parameters are listed in Table \uppercase\expandafter{\romannumeral1}.
\begin{table}[!htp]
\caption{Simulation Parameters}
\vspace{-2em}
\centering{}
\begin{tabular}[t]{|l|l|}
    \hline
    Parameter & Value \tabularnewline
    \hline
    \hline
    Cell radius & 250 m\tabularnewline
    \hline
    Bandwidth, $B$ &10~MHz\tabularnewline
    \hline
    Noise power density, $N_0$ &-174 dBm/Hz \tabularnewline
    \hline
    Path loss exp.  & 4\tabularnewline
    \hline
    Transmission power, $p_{k}$  & 24 dBm\tabularnewline
    \hline
    Raw video size, $L_k$  &[10, 100] Mbits \tabularnewline
    \hline
    Device compression capacity, $V_k^{\text{d}}$ & [0.5, 2] Mbps \tabularnewline
    \hline
    Edge cloud compression capacity, $V^{\text{c}}$  &40 Mbps \tabularnewline
    \hline
\end{tabular}
\end{table}

\subsection{Performance Comparison among Three Models}
We first compare the minimum system delays of local compression, edge cloud compression, and partial compression offloading.
\begin{figure}
	\centering
	\subfigure[System delay with the number of devices.]
    {
		\includegraphics[width=0.6\columnwidth]{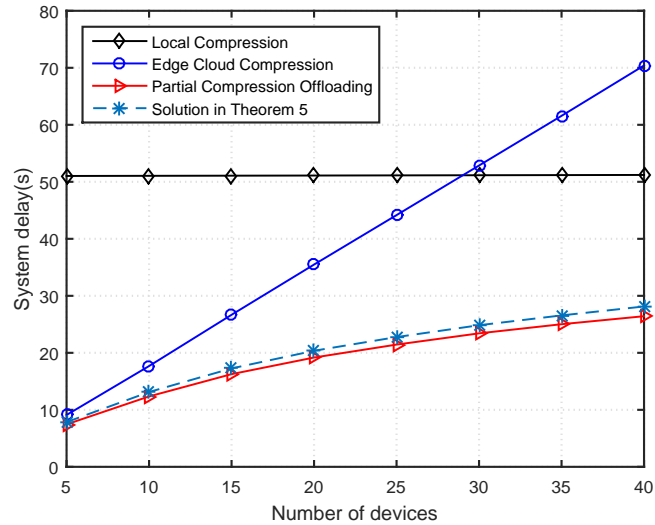}\label{Dsys_K}
	}
	\subfigure[System delay with the device compression capacity.]
    {
		\includegraphics[width=0.6\columnwidth]{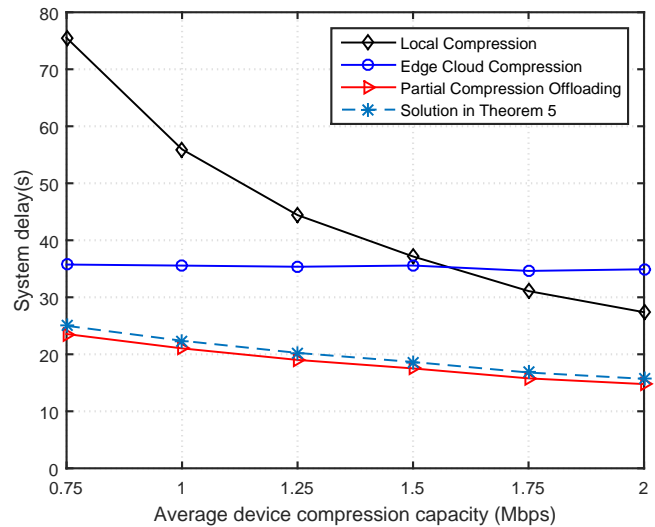}\label{Dsys_Vd}
	}
	\caption{System delay of three models.}\label{System_delay}
\end{figure}

Fig. \ref{Dsys_K} depicts the minimum system delay versus the number of mobile devices in the three different models. First, the system delays of the edge cloud compression and partial compression offloading models increase with the number of mobile devices due to the limited computation resource, while the system delay of the local compression model is approximately invariant since the communication resource is relatively adequate in our simulation. Secondly, by comparing the curves of local compression and edge cloud compression, we can observe that the edge cloud compression performs better than the local compression only when the number of devices is small. The reason can be explained as follows. In case that the number of devices is small, the cloud compression capacity allocated to each device would be larger than the local compression capacity. In this case, it is better to offload computation workload to the edge cloud for compression than local compression from the perspective of delay minimization. On the other hand, as the number of devices grows, the cloud compression capacity allocated to each device would be smaller than the local compression capacity, leading to the better performance of local compression model. Thirdly, the partial compression offloading model has the best performance among the three models since it jointly utilizes the communication and computation resources. The performance gap between the edge cloud compression and partial compression offloading models becomes more evident with the growing number of devices, which indicates that when the number of users becomes large, using the partial compression offloading model can greatly reduce the system delay and improve the QoE for users. Finally, the closed-form solution in Theorem 5 can achieve a near-optimal performance while outperforms both the local compression and edge cloud compression models. It is because under our simulation settings, the compression capacity of mobile devices is much smaller than the corresponding communication capacity. This result demonstrates the effectiveness of our derivation in Theorem 5.

Fig. \ref{Dsys_Vd} shows the minimum system delay versus the average device compression capacity in the three different models. In this simulation, we assume 20 mobile devices in the system while varying the average local compression capacity of all devices from 0.75 Mbps to 2 Mbps. From the figure, we can observe that the system delays of the local compression and partial compression offloading models decrease with the average device compression capacity since both use the local computation resource for video compression. Furthermore, the solution in Theorem 5 has a very close-to-optimal performance especially when the device compression capacity is small, demonstrating its accuracy and applicability in our system.

\subsection{Optimal Resource Allocation in Partial Compression Offloading Model}
Next, we analyze the impact of video size and device compression capacity on the optimal resource allocation in the partial compression offloading model. In this simulation, we assume five devices in the system and keep the video size and local compression capacity of devices 2-5 fixed while varying those parameters of device 1. The detailed simulation parameters for all devices are summarized in Table \uppercase\expandafter{\romannumeral2}.
\begin{table}[!htp]
\caption{Simulation Parameters of Five Devices}
\centering
\vspace{-2em}
\begin{tabular}[t]{|c|c|c|}
	\hline
	Device & Video Size  & Local compression capacity\\
	\hline
	\hline
	1	&10-100 Mbits &0.5-2 Mbps\\
	\hline
	2	&90 Mbits  	&1.2 Mbps\\
	\hline
	3	&80 Mbits	&1.3 Mbps\\
	\hline
	4	&70 Mbits	&1.4 Mbps\\
	\hline
	5	&60 Mbits	&1.5 Mbps\\
    \hline
\end{tabular}
\end{table}
\begin{figure}
	\centering
	\subfigure[Communication resource allocation.]{
		\includegraphics[width=0.6\columnwidth]{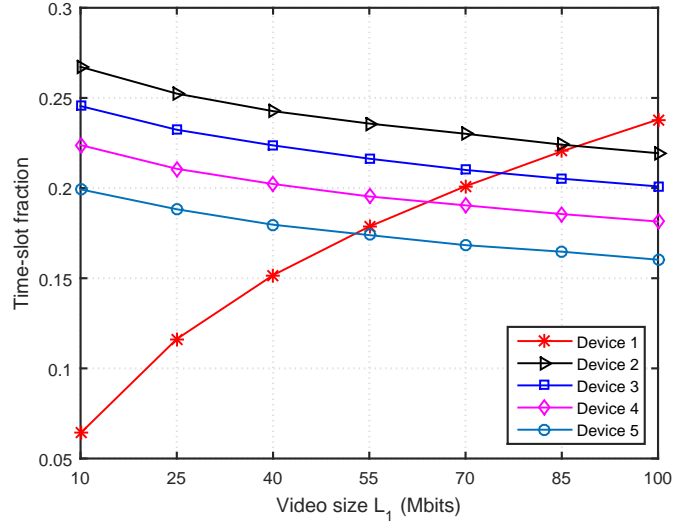} \label{t_L}

	}
	\subfigure[Computation resource allocation.]{
		\includegraphics[width=0.6\columnwidth]{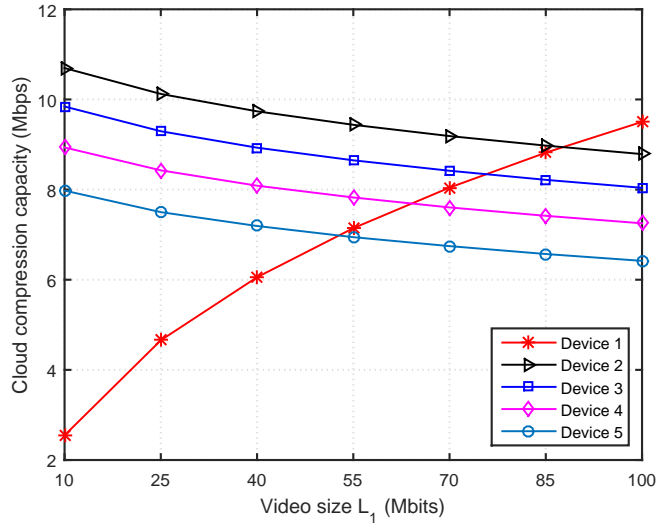} \label{Vc_L}
	}
	\caption{Optimal resource allocation with different video sizes.}\label{Resource_L}
\end{figure}
\begin{figure}
	\centering
	\subfigure[Communication resource allocation.]{
		\includegraphics[width=0.6\columnwidth]{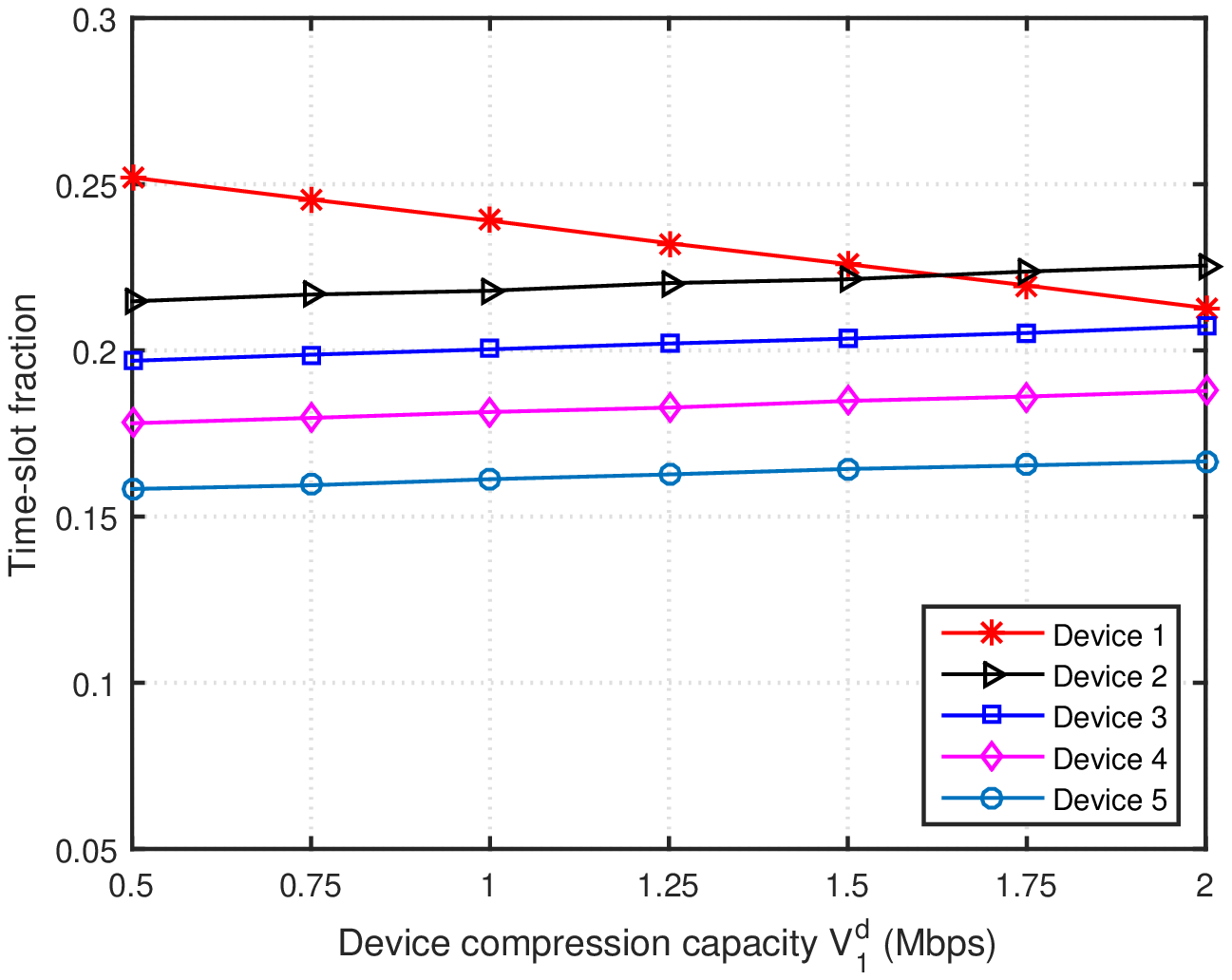}\label{t_Vd}
	}
	\subfigure[Computation resource allocation.]{%
		\includegraphics[width=0.6\columnwidth]{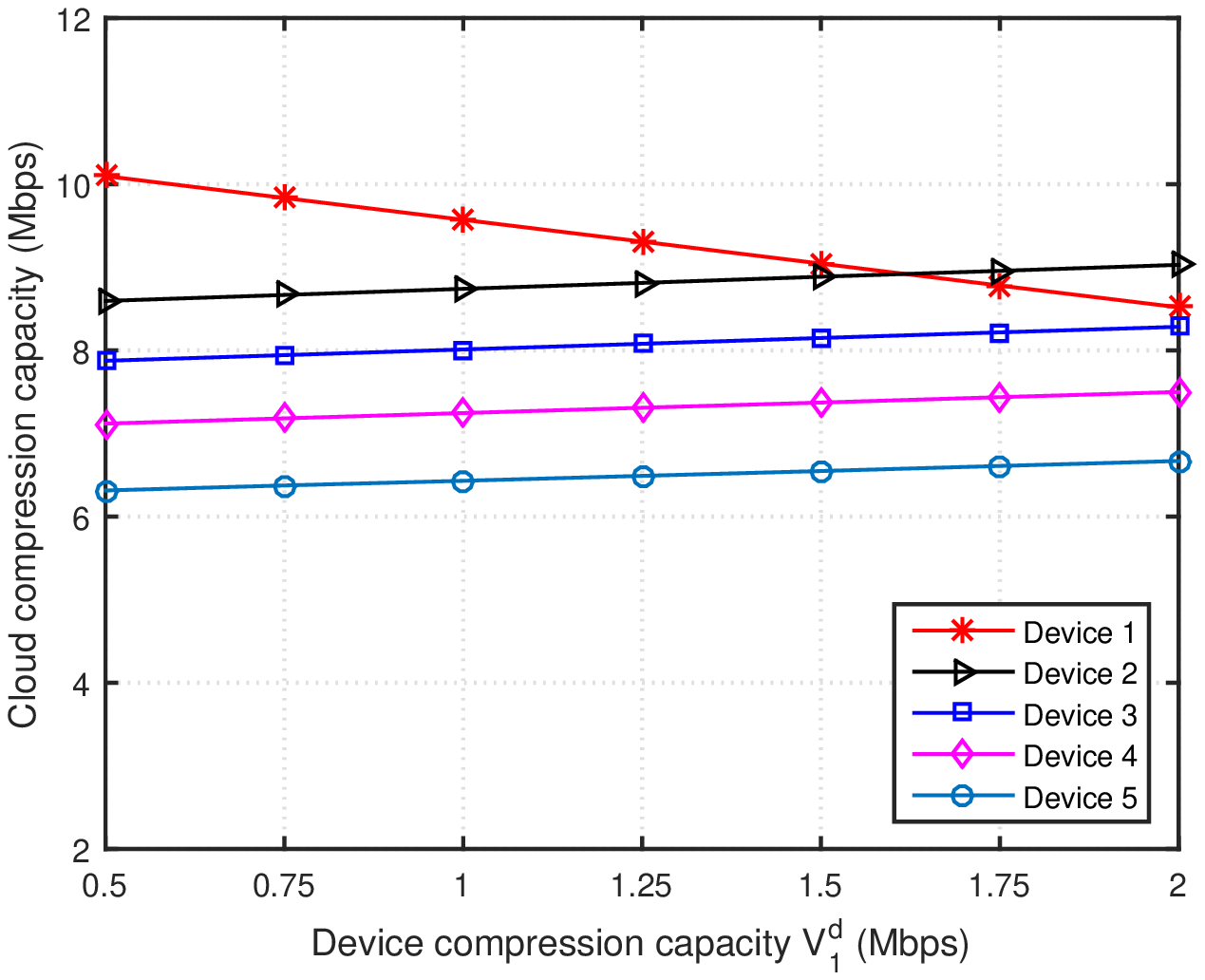}\label{Vc_Vd}
	}
	\caption{Optimal resource allocation with different local compression capacities.}\label{Resource_Vd}
\end{figure}

Fig. \ref{Resource_L} illustrates the optimal time-slot and cloud computation resource allocations with different video sizes of device 1, where the local compression capacity of device 1 is fixed to 1.1     Mbps. It can be observed that the optimal resources $t_1^*$ and $V_1^{\text{c}*}$ allocated to device 1 increase with its video size. On the other aspect, the resources assigned to other devices will consequently decrease. This is rather intuitive due to the fact that more resources should be allocated to device 1 to minimize the system delay as its video size increases. Furthermore, it is shown that the optimal communication and computation resource allocations have almost the same trend, as displayed in Fig. \ref{t_L} and Fig. \ref{Vc_L}. The reason for this outcome is clear since the communication and computation resources have the same effect on computing the end-to-end delay of each device.

Fig. \ref{Resource_Vd} presents the optimal time-slot and edge cloud compression capacity with different local compression capacities of device 1, where the video size of device 1 is fixed to 100 Mbits. It can be observed that the optimal resources $t_1^*$ and $V_1^{\text{c}*}$ allocated to device 1 decrease with its local compression capacity. On the other aspect, the resources allocated to other devices will consequently increase. The reason is that, more resources should be allocated to those devices with lower compression capacity to reduce the weighted-sum delay of all devices. In addition, both optimal communication and computation resource allocations have an approximately linear trend, as shown in Fig. \ref{t_Vd} and Fig. \ref{Vc_Vd}. This is due to the fact that the communication capacity $t_kR_k$, the device compression capacity $V_k^{\text{d}}$, and the cloud compression capacity $V_k^{\text{c}}$ have the same effect on calculating the end-to-end delay of each device.

\section{Conclusion}
This paper investigates joint communication and computation resource allocation for a TDMA-based multi-user MECO system. Our optimization aims to improve the QoE for users by minimizing the weighted-sum delay of all devices. Three models, namely local compression, edge cloud compression, and partial compression offloading, are studied and compared. The optimal solutions for both local compression and edge cloud compression are firstly achieved in closed-form, and some inherent insights are also highlighted. In the partial compression offloading model, we derive the optimal video segmentation strategy for each device in a closed-form expression. Then we formulate a piecewise convex optimization problem, which can be efficiently solved by a developed sub-gradient method. Moreover, to gain some insights, we consider a specific scenario in which communication capacity is much greater than device compression capacity. In this specific scenario, the closed-form solution can be derived. Finally, numerical results demonstrate that the partial compression offloading can efficiently reduce the end-to-end latency as compared against the other two models.

Future works may consider latency-minimization communication and computation resource allocation problem with non-orthogonal channel access where co-channel interference exists. In such a case, our analytical result for the delay performance remains unchanged but the channel capacity expression for each device will be more complicated. Non-convex optimization tools should be utilized to deal with this scenario. Another interesting direction of our future work is to investigate energy-efficiency optimization problem for the multiuser MECO system, i.e., minimizing the overall energy consumption by jointly allocating communication and computation resources.

\begin{appendices}
\section{Proof of Lemma 1}
 To prove Lemma 1, we first calculate a critical case for device $k$ that satisfies
\begin{equation}
	\left\{
    \begin{aligned}
        &D_{\text{comp},k}^{\text{d}} = D_{{\text{tran},k}}^{\text{c}},\\
        &D_{\text{tran},k}^{\text{d}} = D_{{\text{comp},k}}^{\text{c}},
	\end{aligned}
    \right. \label{33}
\end{equation}
which means that the delay for compressing the local part data equals to that for transmitting the edge cloud part data, and the delay for compressing the edge cloud part data also equals to that for transmitting the locally compressed data. Then applying the detailed delay expressions into (\ref{33}) , we can obtain the condition to reach this critical case, as
\begin{equation}
    t_k R_k = \sqrt{\beta V_k^{\text{d}}V_k^{\text{c}}}.\label{34}
\end{equation}
In addition, let us define $\lambda_k^{(1)} = \frac{V_k^{\text{d}}}{t_k R_k+V_k^{\text{d}}}$ which satisfies $D_{\text{comp},k}^{\text{d}} = D_{{\text{tran},k}}^{\text{c}}$ and $\lambda_k^{(2)} = \frac{t_k R_k}{t_k R_k+\beta V_k^{\text{c}}}$ which satisfies $D_{\text{tran},k}^{\text{d}} = D_{{\text{comp},k}}^{\text{c}}$.
Then we can prove Lemma 1 by the following analysis.
\subsection{Case A: $t_k R_k \ge \sqrt{\beta V_k^{\text{d}}V_k^{\text{c}}}$}
In this case, $\lambda_k^{(1)}\le\lambda_k^{(2)}$. When $\lambda_k\in \left[0,\lambda_k^{(1)}\right)$, $D_{{\text{comp},k}}^{\text{d}}< D_{{\text{tran},k}}^{\text{c}}$ and $D_{{\text{tran},k}}^{\text{d}}< D_{{\text{comp},k}}^{\text{c}}$. Therefore, we have $D_k=D_{{\text{tran},k}}^{\text{c}}+D_{{\text{comp},k}}^{\text{c}}=\frac{(1-\lambda_k)L_k}{t_k R_k}+\frac{(1-\lambda_k)L_k}{V_k^{\text{c}}}$, which decreases with $\lambda_k$. When $\lambda_k\in \left[\lambda_k^{(1)},\lambda_k^{(2)}\right]$, $D_{{\text{comp},k}}^{\text{d}} \ge D_{{\text{tran},k}}^{\text{c}}$ and $D_{{\text{tran},k}}^{\text{d}}\le D_{{\text{comp},k}}^{\text{c}}$. Thus we have $D_k = \max (D_{{\text{comp},k}}^{\text{d}}+D_{{\text{tran},k}}^{\text{d}}, D_{{\text{tran},k}}^{\text{c}}+D_{{\text{comp},k}}^{\text{c}})$. Since $D_{{\text{comp},k}}^{\text{d}}+D_{{\text{tran},k}}^{\text{d}}=\frac{\lambda_k L_k}{V_k^{\text{d}}}+\frac{\beta \lambda_k L_k}{t_k R_k}$ increases with $\lambda_k$ while $D_{{\text{tran}}}^{\text{c}}+D_{{\text{comp},k}}^{\text{c}}=\frac{(1-\lambda_k)L_k}{t_k R_k}+\frac{(1-\lambda_k)L_k}{V_k^{\text{c}}}$ decreases with $\lambda_k$, the delay of device $k$ achieves to the minimum when $D_{{\text{comp},k}}^{\text{d}}+D_{{\text{tran},k}}^{\text{d}}=D_{{\text{tran},k}}^{\text{c}}+D_{{\text{comp},k}}^{\text{c}}$, which results in $\lambda_k=\frac{V_k^{\text{d}}\left(t_kR_k+V_k^{\text{c}}\right)}{V_k^{\text{d}}V_k^{\text{c}}\left(1+\beta)+t_kR_k(V_k^{\text{d}}+V_k^{\text{c}}\right)}\in\left[\lambda_k^{(1)},\lambda_k^{(2)}\right]$. Finally when $\lambda_k\in \left(\lambda_k^{(2)},1\right]$, $D_{{\text{comp},k}}^{\text{d}} > D_{{\text{tran},k}}^{\text{c}}$ and $D_{{\text{tran},k}}^{\text{d}} > D_{{\text{comp},k}}^{\text{c}}$. Therefore, we have $D_k = D_{{\text{comp},k}}^{\text{d}}+D_{{\text{tran},k}}^{\text{d}}=\frac{\lambda_k L_k}{V_k^{\text{d}}}+\frac{\beta \lambda_k L_k}{t_k R_k}$, which increases with $\lambda_k$. Based on the above analysis, the optimal video segmentation strategy in this case is $\lambda_k^*=\frac{V_k^{\text{d}}\left(t_kR_k+V_k^{\text{c}}\right)}{V_k^{\text{d}}V_k^{\text{c}}\left(1+\beta)+t_kR_k(V_k^{\text{d}}+V_k^{\text{c}}\right)}$.
\subsection{Case B: $t_k R_k < \sqrt{\beta V_k^{\text{d}}V_k^{\text{c}}}$}
In this case, $\lambda_k^{(1)}>\lambda_k^{(2)}$. When $\lambda_k\in \left[0,\lambda_k^{(2)}\right)$, we have $D_{{\text{comp},k}}^{\text{d}}< D_{{\text{tran},k}}^{\text{c}}$ and $D_{{\text{tran},k}}^{\text{d}}< D_{{\text{comp},k}}^{\text{c}}$. Therefore, we have $D_k=D_{{\text{tran},k}}^{\text{c}}+D_{{\text{comp},k}}^{\text{c}}=\frac{\left(1-\lambda_k\right)L_k}{t_k R_k}+\frac{(1-\lambda_k)L_k}{V_k^{\text{c}}}$, which decreases with $\lambda_k$.
When $\lambda_k\in\left[\lambda_k^{(2)},\lambda_k^{(1)}\right]$, $D_{{\text{comp},k}}^{\text{d}} \le D_{{\text{tran},k}}^{\text{c}}$ and $D_{{\text{tran},k}}^{\text{d}} \ge D_{{\text{comp},k}}^{\text{c}}$. Thus we have $D_k=D_{{\text{tran},k}}^{\text{c}}+D_{{\text{tran},k}}^{\text{d}} = \frac{L_k}{ t_k R_k}\left(1+\left(\beta-1\right)\lambda_k\right)$, which also decreases with $\lambda_k$ because $0<\beta<1$. Finally when $\lambda_k\in\left(\lambda_k^{(1)},1\right]$, $D_{{\text{comp},k}}^{\text{d}} > D_{{\text{tran},k}}^{\text{c}}$ and $D_{{\text{tran},k}}^{\text{d}} > D_{{\text{comp},k}}^{\text{c}}$. Therefore, we have $D_k=D_{{\text{comp},k}}^{\text{d}}+D_{{\text{tran},k}}^{\text{d}}=\frac{\lambda_k L_k}{V_k^{\text{d}}}+\frac{\beta \lambda_k L_k}{t_k R_k}$, which increases with $\lambda_k$. Based on the above discussion, the optimal video segmentation strategy in this case is $\lambda_k^*=\lambda_k^{(1)} =\frac{V_k^{\text{d}}}{V_k^{\text{d}}+t_kR_k}$.

\section{Proof of Theorem 3}
Note that all constraints in Problem 4 are affine. Therefore, Problem 4 is convex if the objective function is convex. In the following, we first prove that $\widehat{D}_k$ is a continuously piecewise convex function.

The Hessian of $\widehat{D}_{k,1}$ is
\begin{equation}
    \bf{H}=
    \left[
        \begin{array}{c c}
            H_{11}	&H_{12}\\
            H_{21}	&H_{22}
        \end{array}
    \right]=
    \left[
        \begin{array}{c c}
        \frac{\partial^2  \widehat{D}_{k,1}}{\partial t_k^2}	                    &\frac{\partial^2  \widehat{D}_{k,1}}{\partial t_k\partial V_k^{\text{c}}}\\
        \frac{\partial^2  \widehat{D}_{k,1}}{\partial V_k^{\text{c}} \partial t_k}	&\frac{\partial^2  \widehat{D}_{k,1}}{\partial(V_k^{\text{c}})^2}
        \end{array}
    \right]. \label{35}
\end{equation}
We can prove that the Hessian in (\ref{35}) is positive-definite by proving all the leading principal mirrors of $\bf{H}$ are positive, as
\begin{equation}
    \begin{aligned}
	\Delta_1=H_{11} = 2L_k
    & \left( \frac{ \beta (1+\beta)\left(V_k^{\text{d}} V_k^{\text{c}}\right)^2 \left( (1+\beta) V_k^{\text{d}} V_k^{\text{c}} + 3 t_k R_k \left( V_k^{\text{d}}+V_k^{\text{c}}
      \right)\right)}{t_k^3 R_k\left(\left(1+\beta\right)V_k^{\text{d}}V_k^{\text{c}} + t_k R_k \left(V_k^{\text{d}}+V_k^{\text{c}}\right)\right)^3}+\right.\\
    &~~~\left.\frac{ \left(t_k R_k\right)^2 \left( V_k^{\text{d}} + V_k^{\text{c}} \right) \left( 3\beta V_k^{\text{d}} V_k^{\text{c}} \left( V_k^{\text{d}} + V_k^{\text{c}}\right) + t_k R_k \left(\beta
        \left(V_k^{\text{d}}\right)^2+ \left(V_k^{\text{c}}\right)^2 \right) \right)} {t_k^3 R_k\left(\left(1+\beta\right)V_k^{\text{d}}V_k^{\text{c}} + t_k R_k \left(V_k^{\text{d}}+V_k^{\text{c}}\right)\right)^3} \right)>0,
    \end{aligned} \label{36}
\end{equation}
\begin{equation}
	\begin{aligned}
	\Delta_2
    &= H_{11}H_{22}-H_{12}H_{21}\\
    &= 4L_k^2 \left( t_k R_k + \beta V^d_k \right)^2 \left( \frac{ \beta (1+\beta) \left(V_k^{\text{d}}\right)^2 V_k^{\text{c}} \left( (1+\beta) V_k^{\text{d}} V_k^{\text{c}} + t_k R_k  \left(
       2V_k^{\text{d}}+3V_k^{\text{c}}\right) \right)} {t_k^3 R_k \left( (1+\beta) V_k^{\text{d}} V_k^{\text{c}} + t_k R_k \left( V_k^{\text{d}} + V_k^{\text{c}}\right) \right)^5}+\right.\\
    &~~~~\left.\frac{ \left(t_k R_k\right)^2  \left(  t_k R_k \left( \beta \left(V_k^{\text{d}}\right)^2 + \left(V_k^{\text{c}}\right)^2 \right) + \beta V_k^{\text{d}} \left( \left(V_k^{\text{d}}\right)^2 + 4V_k^{\text{d}} V_k^{\text{c}} + 3\left(V_k^{\text{c}}\right)^2   \right)\right)}   {t_k^3 R_k \left( (1+\beta) V_k^{\text{d}} V_k^{\text{c}} + t_k R_k \left( V_k^{\text{d}} + V_k^{\text{c}}\right) \right)^5} \right)>0.
	\end{aligned} \label{37}
\end{equation}
Therefore, $\widehat{D}_{k,1}$ is strictly convex on both $t_k$ and $V_k^{\text{c}}$.

Next, we prove that $\widehat{D}_{k,2}$ is also convex on $t_k$ and $V_k^{\text{c}}$. The second-order partial derivative of $\widehat{D}_{k,2}$ on $t_k$ fulfills
\begin{equation}
	\frac{\partial ^2 \widehat{D}_{k,2}}{\partial t_k^2} = \frac{2L_k\left(\left(t_k R_k\right)^3 + 3 \beta V_k^{\text{d}}\left(t_k R_k\right)^2 + 3\beta
    \left(V_k^{\text{d}}\right)^2 t_k R_k + \beta \left(V_k^{\text{d}}\right)^3\right)}{t_k^3 R_k \left(t_k R_k + V_k^{\text{d}}\right)^3}>0. \label{38}
\end{equation}
Since $\widehat{D}_{k,2}$ does not change over $V_k^{\text{c}}$, $\widehat{D}_{k,2}$ is convex on $t_k$ and $V_k^{\text{c}}$. Moreover, it can be easily verified that $\widehat{D}_{k,1}=\widehat{D}_{k,2}$, $\dfrac{\partial \widehat{D}_{k,1}}{\partial t_k}\neq \dfrac{\partial \widehat{D}_{k,2}}{\partial t_k}$, and $\dfrac{\partial \widehat{D}_{k,1}}{\partial V_k^{\text{c}}}\neq \dfrac{\partial \widehat{D}_{k,2}}{\partial V_k^{\text{c}}}$ at $t_k R_k=\sqrt{\beta V_k^{\text{d}}V_k^{\text{c}}}$. Therefore $\widehat{D}_k$ is a continuous and piecewise function. Then computing the difference between $\widehat{D}_{k,1}$ and $\widehat{D}_{k,2}$, we have
\begin{equation}
\begin{aligned}
    \widehat{D}_{k,1}-\widehat{D}_{k,2} &= \frac{L_k}{t_k R_k}\frac{\left(t_k R_k+\beta V_k^{\text{d}} \right) \left(\left(t_k R_k\right)^2-\beta V_k^{\text{d}} V_k^{\text{c}}\right)} {\left(V_k^{\text{d}}
                                            V_k^{\text{c}}(1+\beta)+t_k R_k \left(V_k^{\text{d}}+V_k^{\text{c}}\right) \right)\left(t_k R_k+V_k^{\text{d}}\right)}. \label{41}
\end{aligned}
\end{equation}
Therefore, when $t_k R_k\ge\sqrt{\beta V_k^{\text{d}}V_k^{\text{c}}}$, $\widehat{D}_{k,1} \ge \widehat{D}_{k,2}$, otherwise $\widehat{D}_{k,1} < \widehat{D}_{k,2}$. According to (\ref{23}), $\widehat{D}_k$ can be rewritten as $\max \{\widehat{D}_{k,1}, \widehat{D}_{k,2}\}$, which is convex on $t_k$ and $V_k^{\text{c}}$ since the pointwise maximum preserves convexity. Furthermore, the objective function $\sum_{k=1}^K \alpha_k \widehat{D}_k$ is the summation of a set of convex functions, which is also convex. This ends the proof.

\section{Proof of Theorem 4}
In the following, we will prove that the iterative method in Theorem 4 could converge to the optimal resource allocation solution, denoted as ${\bm{x}}^*$. First, the Euclidean distance between the $(n+1)^{\text{th}}$ iteration solution and the optimal solution can be calculated as
\begin{align}
    \left|\left|{\bm{x}}^{(n+1)}-{\bm{x}}^*\right|\right|_2^2
        &= \left|\left|{\bm{x}}^{(n)}-\phi_n{\bm{g}}^{(n)}-{\bm{x}}^*\right|\right|_2^2  \label{converge_1}\\
        &= \left|\left|{\bm{x}}^{(n)}-{\bm{x}}^*\right|\right|_2^2-2\phi_n{\bm{g}}^{(n)T}\left({\bm{x}}^{(n)}-{\bm{x}}^*\right)+\phi_n^2\left|\left|{\bm{g}}^{(n)}\right|\right|_2^2  \label{converge_2}\\
        &\le\left|\left|{\bm{x}}^{(n)}-{\bm{x}}^*\right|\right|_2^2-2\phi_n\left(F({\bm{x}}^{(n)})-F^*\right)+\phi_n^2\left|\left|{\bm{g}}^{(n)}\right|\right|_2^2 \label{converge_3}\\
        &\le\left|\left|{\bm{x}}^{(0)}-{\bm{x}}^*\right|\right|_2^2-2\sum_{i=0}^{n}\phi_i\left(F({\bm{x}}^{(i)})-F({\bm{x}}^*)\right)+\sum_{i=0}^{n}\phi_i^2\left|\left|{\bm{g}}^{(i)}\right|\right|_2^2, \label{converge_4}
\end{align}
where $\left|\left|{\bm{x}}\right|\right|_2$ is the Euclidean norm of ${\bm{x}}$ and $\bm{g}^{(n)T}$ represents the transpose of $\bm{g}^{(n)}$. Notice that the  inequality operation (\ref{converge_3}) is based on the convexity of $\sum_{k=1}^{K}\alpha_k \widehat{D}_k$, as $F({\bm{x}}^*)\ge F({\bm{x}}^{(n)})+{\bm{g}}^{(n)T}\left({\bm{x}}^*-{\bm{x}}^{(n)}\right)$.

Let us denote $F^{(n)}_{\text{best}}=\min_{i=0}^n F\left({\bm{x}^{(i)}}\right)$. Then we have
\begin{equation}
    \sum_{i=0}^{n}\phi_i\left(F\left({\bm{x}}^{(i)}\right)-F\left({\bm{x}}^*\right)\right)\ge \left(F^{(n)}_{\text{best}}-F\left({\bm{x}}^*\right)\right)\left(\sum_{i=0}^{n}\phi_i\right).\label{44}
\end{equation}
Substituting (\ref{44}) into (\ref{converge_4}), we can derive the upper bound difference between $F^{(n)}_{\text{best}}$ and $F(\bm{x}^*)$, as
\begin{align}
    F^{(n)}_{\text{best}}-F\left(\bm{x}^*\right)
    & \le \frac{\left|\left|{\bm{x}}^{(0)}-{\bm{x}}^*\right|\right|_2^2-\left|\left|{\bm{x}}^{(n+1)}-{\bm{x}}^*\right|\right|_2^2 + \sum_{i=0}^{n}\phi_i^2\left|\left|{\bm{g}}^{(i)}\right|\right|_2^2}
          {2\sum_{i=0}^{n}\phi_i}  \label{45}\\
    & \le \frac{\left|\left|{\bm{x}}^{(0)}-{\bm{x}}^*\right|\right|_2^2-\left|\left|{\bm{x}}^{(n+1)}-{\bm{x}}^*\right|\right|_2^2+G\sum_{i=0}^{n}\phi_i^2}{2\sum_{i=0}^{n}\phi_i},\label{46}
\end{align}
where $G=\max_{i=0}^n \left|\left|{\bm{g}}^{(i)}\right|\right|_2^2$.

Under such circumstances, if we select $\phi_n$ that satisfies $\sum_{n=0}^{\infty}\phi_n=\infty$ and $\sum_{n=0}^{\infty} \phi_n^2<\infty$, such as $\phi_n=\dfrac{1}{n+1}$, $F^{(n)}_{\text{best}}-F({\bm{x}}^*)$ will gradually converge to zero. Moreover, to accelerate the convergence speed, we can select the Polyak step size $\phi_n=\dfrac{F\left({\bm{x}}^{(n)}\right)-F^{(n)}_{\text{best}} + \gamma_n}{\left|\left|{\bm{g}}^{(n)}\right|\right|_2^2}$, where $\gamma_n$ satisfies $\sum_{n=0}^{\infty}\gamma_n=\infty$ and $\sum_{n=0}^{\infty}\gamma_n^2<\infty$ \cite{Convex_Optimization_II}. Then the iteration will linearly converge to the optimal solution $\bm{x}^*$. This ends the proof.

\section{Proof of Theorem 5}
It can be easily verified that the end-to-end delay expression (\ref{31}) is strictly convex on $t_k$ and $V_k^{\text{c}}$ using the derivation method in Appendix B. Therefore, we can utilize the KKT conditions to derive the closed-form solution. Let $\{\overline{t}_k^*, \overline{V}_k^{\text{c}*}\}$ denote the optimal solution for this specific scenario. Then the Lagrange function of $\sum_{k=1}^K \alpha_k \overline{D}_k$  can be expressed as
\begin{equation}
    \overline{L}_{\text{P}} = \sum_{k=1}^K \alpha_k \overline{D}_k + \theta \left(\sum_{k=1}^K t_k - 1\right) + \omega \left(\sum_{k=1}^K V_k^{\text{c}} - V^{\text{c}}\right). \label{47}
\end{equation}

Applying the KKT conditions leads to the following necessary and sufficient conditions
\begin{equation}
    \frac{\partial \overline{L}_{\text{P}}}{\partial \overline{t}_k^*} =
        -\frac{\alpha_k L_k R_k \left(\overline{V}_k^{\text{c}*}\right)^2 }{ \left( \overline{t}_k^* R_k  V_k^{\text{d}} + \overline{t}_k^* R_k \overline{V}_k^{\text{c}*} + V_k^{\text{d}} \overline{V}_k^{\text{c}*} \right)^2} +\theta^*
    \left\{
    \begin{aligned}
        &>0,~~\overline{t}_k^*=0,\\
        &=0,~~\overline{t}_k^*>0,
    \end{aligned}
    \right.\label{48}
\end{equation}
\begin{equation}
    \frac{\partial \overline{L}_{\text{P}}}{\partial \overline{V}_k^{\text{c}*}}=
        -\frac{\alpha_k L_k \left(\overline{t}_k^* R_k \right)^2 }{ \left( \overline{t}_k^* R_k  V_k^{\text{d}} + \overline{t}_k^* R_k \overline{V}_k^{\text{c}*}+V_k^{\text{d}}\overline{V}_k^{\text{c}*} \right)^2} + \omega^*
    \left\{
    \begin{aligned}
        &>0,~~\overline{V}_k^{\text{c}*}=0,\\
        &=0,~~\overline{V}_k^{\text{c}*}>0,
    \end{aligned}
    \right. \label{49}
\end{equation}
\begin{equation}
    \theta^*\left(\sum_{k=1}^K \overline{t}_k^* - 1\right)=0,~\sum_{k=1}^K \overline{t}_k^* \le 1,~\theta^*\ge 0,\label{50}
\end{equation}
\begin{equation}
    \omega^*\left(\sum_{k=1}^K \overline{V}_k^{\text{c}*}- V^{\text{c}}\right)=0,~\sum_{k=1}^K \overline{V}_k^{\text{c}*}\le V^{\text{c}},~\omega^*\ge 0.\label{51}
\end{equation}
Based on the above conditions, we can derive the optimal resource allocation solution for the special case, as
\begin{equation}
	\left\{
    \begin{aligned}
        &\overline{t}_k^* = \frac{\overline{V}_k^{\text{c}*}\left(\sqrt{\frac{\alpha_k L_k R_k}{\theta^*}}-V_k^{\text{d}} \right)^+}{R_k \left( V_k^{\text{d}} + \overline{V}_k^{\text{c}*} \right)},~\forall
          k \in \mathcal{K},\\
        &\overline{V}_k^{\text{c}*} = \frac{\overline{t}_k^* R_k \left(\sqrt{\frac{\alpha_k L_k}{\omega^*}}-V_k^{\text{d}} \right)^+}{\overline{t}_k^*R_k + V_k^{\text{d}}},~\forall k \in \mathcal{K}.
	\end{aligned}
    \right. \label{52}
\end{equation}
This ends the proof.
\end{appendices}

\end{document}